%% file: sensorchain.tex
\documentclass[journal]{IEEEtran}
\usepackage{cite}

\ifCLASSINFOpdf
  \usepackage[pdftex]{graphicx}
\else
  \usepackage[dvips]{graphicx}
\fi
\usepackage{url}

\usepackage{plantuml}

\begin{document}

\title{Security in Distributed Systems by Verifiable Location-Based Identities}

\author{Simon Tschirner, Katharina Zeuch, Sascha Kaven, Lorenz Bornholdt, Volker Skwarek  \\
firstname.surname@haw-hamburg.de\\
\textit{RTC Digital Business Processes, Faculty of Life Sciences}\\
Hamburg University of Applied Sciences, Ulmenliet 20, 21033 Hamburg, Germany \\ 
corresponding author: simon.tschirner@haw-hamburg.de
}

\maketitle

\begin{abstract}
Proof-of-Location (PoL) is a lightweight security concept for Internet-of-Things (IoT) networks, focusing on the sensor nodes as the least performant and most vulnerable parts of IoT networks. PoL builds on the identification of network participants based on their physical location. It introduces a secondary message type to exchange location information. Via these messages, the nodes can verify the integrity of other network participants and reach a consensus to identify potential attackers and prevent malicious information from spreading. 

The paper presents the concretization of the concept to allow implementation on real hardware. The evaluation based on this implementation demonstrates the feasibility of PoL and enables identifying further steps to develop a deployable protocol. 
\end{abstract}

\begin{IEEEkeywords}
Wireless sensor networks, Internet-of-Things, Consensus, Security, Trust.
\end{IEEEkeywords}

\IEEEpeerreviewmaketitle

\input{sections/1-intro.tex}
\input{sections/2-basics.tex}
\input{sections/3-security.tex}

\input{sections/4-concept.tex}
\input{sections/5-architecture.tex}

\input{sections/6-implementation.tex}

\input{sections/7-conclusion.tex}

\appendices

\section*{Acknowledgment}
This research was performed within the project Watergridsense
4.0 and supported by the German Ministry of Education
and Research (BMBF Grant No. 02WIK1474B).

\ifCLASSOPTIONcaptionsoff
  \newpage
\fi



\bibliographystyle{IEEEtran}

%

%








\end{document}

%% file: sections/1-intro.tex
\section{Introduction} \label{chap-intro}

\IEEEPARstart{T}{he} Internet-of-Things (IoT) belongs to the fastest developing fields during the past decade. With its growth, it is safe to assume that there are (soon) tens of billions of IoT-devices existing \cite{statista_2016}. Their application spread to almost every field of life, covering everything from industrial production plants to children toys, from large scale sensor applications for environmental monitoring to health care applications, from automotive devices to home appliances. Among these applications are many safety-critical tasks, where a malfunction or malicious tampering with devices can impact human health, may have a large financial impact, or tampers with privacy.

Their increasing spread and importance also increase the interest attackers are developing for IoT-devices. 

At the same time, IoT-networks are prone to attacks, due to their typical properties. Alaba et al. \cite{Alaba.2017} provide a comprehensive review on IoT-security. Naturally, inter-connectedness combined with little resources at their disposal characterises IoT-devices. Requirements on costs, size and energy consumption are strong, which means that their design provides computation and memory capacity close to the bare minimum required to fulfil their tasks.

Traditional security mechanisms, e.g. encryption of memory and data transmission, in turn requiring strong computational power, contradicting the common light-weight nature of IoT-devices \cite{hameed_security_2019}.

An aspect of IoT security is the ability to trust in data received from a device that is part of the network. This use case is typical for wireless sensor networks (WSN), which are, when connected via a gateway to the internet, also a part of the IoT. In large WSNs, sensor nodes (SN) cover an area, exceeding the transmission range of a single SN. These scenarios use multi-hop communication, where multiple SNs forward data from its source to the sink (gateway). The WSN has to ensure that sensed data will not be altered (in an undesired way) on its path from the original SN to the sink.

Thus, IoT requires security mechanisms despite containing SNs as the least performant parts. Recent surveys identified many security issues in IoT \cite{hameed_security_2019, meneghello_iot_2019}, while approaches to secure IoT have been proposed in the literature as well, e.g. \cite{farooq_efficient_2019}.

One lightweight security approach focused on WSNs is provided by \cite{bornholdt2019proof}, a so-called \emph{Proof-of-Location} (PoL). They propose to secure a WSN by using a trust mechanism consisting of a combination of redundant localisation of sensor nodes and consensus generation within a WSN.

The main purpose of this paper is to address the challenge arising from a combination of the light-weight nature of IoT devices with the computational complexity of effective security mechanisms. Thereby, the following research questions will be addressed:
\begin{description}
    \item[Q1.] How can PoL become an implementable, usable protocol?
    \item[Q2.] Can PoL serve as a short-range communication light-weight security mechanism?
\end{description}

This paper further defines the PoL protocol, reaching an implementable state. Detailed extensions for the concept of PoL include its implementation and evaluation in the field of short-range communication. The presented extensions incorporate the understanding of an IoT device's identity proposed by Wöhnert et al. \cite{wohnert2020secure}.

The rest of the paper is structured as follows: First, the background focuses on the usage of identity to create trust in data from IoT devices. Further identity-based attacks, avoidable by using the presented concept, are depicted. Section \ref{chap-security} describes how the presented concept has been developed to create a solution to the identified identity-based attacks. The description of the concept itself follows in Section \ref{chap-concept}. Implementation details are shown in Section \ref{chap-architecture} and experimental results from a concrete implementation are presented in Section \ref{chap-implementation}. Finally, conclusions and future work are summarised.

%% file: sections/2-basics.tex
\section{Trust and Identities in the Internet of Things} \label{chap-basics}

Securing a WSN contains two aspects: 
\begin{itemize}
    \item ensuring trust in the system's data and
    \item ensuring that only authorised devices enter the network.
\end{itemize}

First, a basic understanding of the term \emph{trust} is necessary. Often, trust is defined as requiring psychological and sociological abilities \cite{schultz_trust_2006}. Considering that WSNs do not offer these abilities, a more technical approach is needed. Taking into account \cite{flowerday_trust_2006} and \cite{josang_right_nodate}, trust in a computer(-system) and its data resembles the fact that the computer(-system), device or algorithm was not manipulated by a malicious agent \cite{josang_right_nodate}. This absence of manipulation leads to a reduction of uncertainty about data and devices within the network. According to \cite{flowerday_trust_2006}, this is a necessary prerequisite for the development of trust. Consequently, this paper assumes that trust in the system's devices implies trust in the system's data.

Second, the term \emph{identity} has to be discussed. ISO defines identity as ``characteristics determining who or what a person or thing is'' \cite{ISO-identity_23093}. A unique identity gives us a mean to distinguish every single device from any other device in a network.

A closer analysis of possible attacks on WSNs reveals the possibility to reduce several different attacks to the ability to pretend false identities. Using this ability is known as Sybil attack, first mentioned by \cite{douceur2002sybil}. Examples of related attacks are a replay attack as well as an active man-in-the-middle attack. Both aim to inject false data or commands into a network. Also, Denial-of-Service (DoS) attacks, working with any form of flooding mechanisms, have to present a valid identity to make SNs process messages instead of immediately dismiss them. 
Unique device identification is required to avoid unauthorised devices entering the network,

Wöhnert \cite{wohnert2020secure} propose an approach for defining the identity of cyber-physical IoT devices. They define identity as a combination of a finite number of attributes that the IoT device has. The concept presented in this paper takes advantage of the distributed nature of WSN to secure such identities of the network's devices. It uses mechanics inspired by distributed ledger technology (DLT) -- consensus and data distribution.

The consensus is used in DLT-systems to aim for a system wherein the trustor does not need to trust one trustee or a third party. Instead, it has to trust in the system itself, its algorithm and the totality of entities. Often, consensus mechanisms implement Byzantine Fault Tolerant protocols as described by Lamport \cite{lamportByzantineGeneralsProblem1982a}. The other DLT-principle is the distribution of data. Every entity can own a copy of the ledger, including all network transactions, to ensure transparency. Later this paper describes how the proposed approach uniquely identifies devices in the network, using distributed, identity-related data in combination with a consensus on this data.

Given the proposed mechanism of secured, unique identification, it is even possible to estate trust in data: assuming that a device is trustworthy, based on its identity and the past behaviour of this identity, data provided by this identified device is trustworthy.

%% file: sections/3-security.tex
\section{Securing WSNs via Location-Based Consensus}
\label{chap-security}

As described in the previous section, a set of entities can build a unique identity. This section defines these entities as well as the DLT-mechanisms that secure the identity-related information. 

SNs can have various attributes that create their identity. The more attributes used for identification, the easier it is to guarantee a unique identification. However, the usage of complex identity profiles in WSN is inadequate due to memory and computational restrictions. Therefore, a more light-weight approach is needed. Several approaches include the location of a node, e.g. as a context for measured data \cite[p.\ 2519]{oscar_wang_artsense:_2013}. Its location is a unique attribute of a device, as there is physically no possibility that two devices are at the same location. Thus location would be a unique attribute, forming a device's identity. Using a node's location for identification and trust in its data has been introduced by \cite{skwarek2017blockchains} and further described as \emph{Proof-of-Location} (PoL) in \cite{bornholdt2019proof}. This paper focuses on a detailed description of PoL and presents concrete steps towards its implementation.
A focus lies on the DLT-mechanisms that should prohibit or at least identify attempts to tamper with data integrity as stated by \cite[chapter ``paying for integrity''] {drescherBlockchainBasicsNonTechnical2017}. 

PoL, as described by \cite{bornholdt2019proof}, is a very light-weight approach, as the required authentication information is naturally available in WSNs and often already accessible through the used hardware, e.g. via the received signal strength indicator (RSSI). This value gives a rough estimation of the distance between two nodes. PoL uses the lateration of RSSI-values from different nodes in range to derive a node's location. Despite knowing that using RSSI for localisation has its limitations \cite{RSSI-limitations}, it will be used as an example throughout this paper, exploring if the proposed concept around PoL works in principle. In practice, depending on the application, one has to expect that the resolution of positioning based on RSSI is not high enough to guarantee consistency and uniqueness of a derived SN's location.

Trust in an identity based on the location, as proposed by PoL, can be categorised into direct trust and recommended trust \cite[p.\ 46]{li_trust_2007}. On the one hand, there exists direct trust, SNs have in the node they localise themselves. On the other hand, parties not involved in the localisation must have confidence in the localised node due to tamper-proof determination of its location and integrity of the result.
Consequently, trust within a system builds upon the identification of the nodes based on their location.

Additionally, the consensus among the network participants, the sensor nodes, on data integrity is key for their persistence. When a certain number of nodes in the network with (explicit) or without (implicit) communication agrees on the trustworthiness of certain information, it gets stored by the system (without further defining or explaining different implementations of the storage right now).

From a security perspective, implementation of PoL covers all aspects of the CIA\textsuperscript{3}-goals \cite{stallingsComputerSecurityPrinciples2018} but confidentiality, which still has to be achieved using standard cryptographic methods. All other aspects are covered by a combination of a BFT communication protocol and an identity check (see Section \ref{chap-concept}). Data \emph{integrity} by means of unauthorised alteration according to ISO-definitions such as \cite{internationalstandardisationorganisationISOIEC2918222013} or at least its detection is achieved by the BFT \cite{lamportByzantineGeneralsProblem1982a} itself: The combination of redundant communication and repetition using different message propagation paths helps to detect unintended alterations. Note that this method does not secure that the data itself is correct. As the data originates from a sensor, the sensor's output must be true and valid. \emph{Availability} is defined as accessibility and usability of the data on demand \cite{internationalstandardisationorganisationISOIEC2918222013}. Technically seen, the availability is ensured by the broadcast of the different message types further described in Section \ref{chap-concept}. Therefore, messages are not sent via one single, attackable way but to many recipients. \emph{Authenticity} (= verifiable identity of the origin of messages, derived from \cite{internationalstandardisationorganisationISOIEC2918222013}) and \emph{accountability} are created by independent verification of the identity of the original sender of a message using distance information of multiple recipients. They have to additionally exchange their knowledge about the distance according to BFT-communication, as explained in Section \ref{chap-concept}.

%% file: sections/4-concept.tex
\section{Concept of Proof-of-Location} \label{chap-concept}

A conclusion from the discussion of trust and identity from the previous sections is that network communication can be secured by identification of \emph{trustworthy} and \emph{untrustworthy} communication participants. PoL utilises this principle and uses a node's location as the key component for its identity. The most suitable scenarios for PoL are those where nodes are typically in a fixed location or at least in a fixed relative location to each other. Examples are (long-term) measurement campaigns where nodes are supposed to report certain information from determined locations or logistics scenarios, where nodes are located inside a freight container. The container moves, while the relative location of the nodes inside stays the same.  For a static network, nodes not changing their location are defined as \emph{trustworthy}. Nodes changing their location are seen as \emph{untrustworthy} and such is the information spread by those. 

Considering mobile nodes requires additional mechanics. Either a node has to report to the network when it becomes aware of being moved, e.g.\ using values from acceleration sensors, or the protocol needs to be adjusted. A possibility is that the mobility of a node becomes part of its identity. Another option is the usage of additional attributes or knowledge of the network's history. However, these considerations are beyond the scope of the presented work.

\subsection{Basic Principle}

A precondition is that each node knows its exact, relative location, e.g.\ as a result of an initialisation phase (not considered in this paper). During further communication, nodes have to create a representation of the network topology by identifying and storing other nodes' positions additional to attributes, which can be considered part of their identity.  Once a node has an overview of the surrounding network, it is possible to relate received messages, their source and the source's stored location. It is possible to verify the source's location and thus the trustworthiness of its transmitted information. In consequence, nodes can monitor manipulation attempts. 

The topology storage plays a central role in the implementation of PoL. It stores the different node's locations, related measurement data and further attributes needed for their identification. Thus it is a node's memory of the network and its main tool to identify attacks.

\subsection{Message Types}

The implementation of PoL presented in this paper utilises three message types for communication inside the network: \emph{payload messages}, \emph{BFT messages}, and \emph{alert messages}. Payload messages contain the actual payload, i.e.\ the sensed data. BFT messages transmit and exchange location-related information, a precondition to identifying (malicious) nodes. Alert messages indicate that a node detects a problem,  either a violation of a predefined condition connected to the original measurement task or a manipulation attempt on the network or identities inside the network.

Since only the first message type spreads sensed information, messages of the other two types (except alerts connected to the measurement task) introduce additional overhead caused by the implementation of PoL. Thus, to keep the approach light-weight, usage of these types has to be limited.

\subsubsection{Payload Message}

\begin{figure}[ht]
  \centering
  \includegraphics[width=0.8\linewidth]{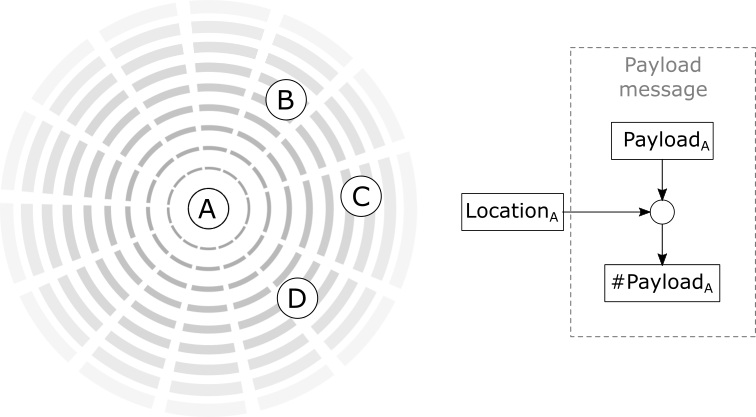}
  \caption{Payload messages (here emitted by A) include payload and payload signed with the node's location. Receiving nodes can use their information about the sender's location to confirm that both sender and receiver share the same information about the senders' position.}
  \label{fig:paymessage}
\end{figure}

Payload messages contain sensed data. This message includes the payload and the payload signed with the node's location (cf.\ Figure \ref{fig:paymessage}). Signing the payload, on the one hand, connects it to a particular location.  On the other hand, it allows the receivers (B, C and D) to validate the sender's (A) identity.

When such a payload message is received, the recipients will try to verify the sender's location. They could, e.g., sign the original payload again, using the sender's location from their topology storage. The transmission contains additional contextual information (e.g. has the sender previously sent the same class of information from the same position?). Extraction of this and comparison to historical information allows for further analysis of a sender's and message's integrity. 

\subsubsection{BFT Message}

BFT messages spread localisation information inside the network. In the current implementation, the \emph{topology storage} is a kind of memorised map in which each node stores the identity information it gains about other nodes. Identity information about a node can be sent by the node itself, extracted or measured from direct communication with that node, or information about a node, spread by a third node. 

RSSI values are a measured value from communication, as further explained in the following sections. BFT messages are the other two sources of information. In the protocol presented in this paper, BFT messages are the first level of distrust or dissent: Whenever a node receives a message from a sender of which it cannot prove identity, it emits a BFT message. In this case, an identity that cannot be proven means that not enough information is present in the receiving node's topology storage or that measured information, e.g.\ the RSSI value, is not consistent with the available information in topology storage. 

\begin{figure}[h!]
  \centering
  \includegraphics[width=0.8\linewidth]{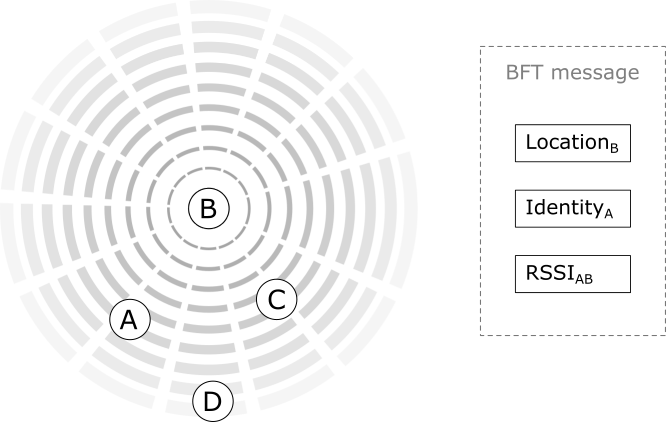}
  \caption{B is emitting a BFT message, which expresses dissent with A's identity and typically contains its sender's (B) location, stored identity of the payload message sender (A), and measured $RSSI_{AB}$.}
  \label{fig:bftmessage}
\end{figure}

BFT messages (cf.\ Figure \ref{fig:bftmessage}) contain the BFT message sender's (B) location, stored identity of the payload message sender (A), and measured $RSSI_{AB}$. Receivers can add this information (if trusted) to their topology storage to complete their picture of the network. The node, which is the object of the BFT message, e.g.\ the identity of which is challenged, can respond, too. BFT messages can be seen as necessary to create a consensus -- or rather dissent -- as the more BFT messages from different nodes are emitted about the same node (object of the BFT message), the clearer it becomes that the object actually might be manipulated.

\subsubsection{Alert Message}

\begin{figure}
  \centering
  \includegraphics[width=0.8\linewidth]{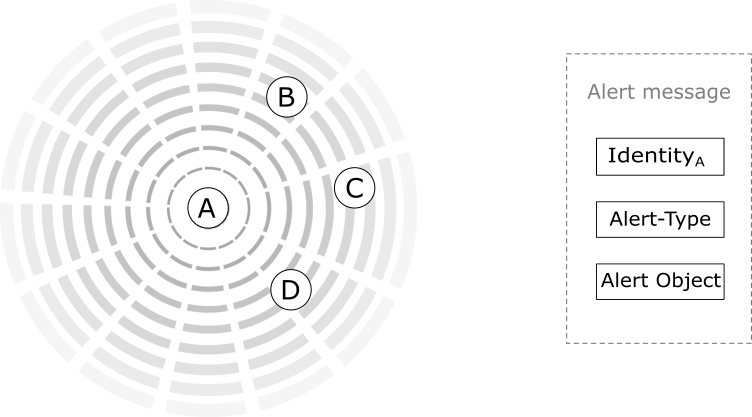}
  \caption{A is emitting an alert message, containing its identity, the alert-type and object.}
  \label{fig:alertmessage}
\end{figure}

The paper distinguishes two completely different types of alerts: Alerts according to the measurement tasks, e.g.\ a measured temperature reaches a set limit, and alerts regarding network infrastructure. Both types are semantically similar, high priority messages, required to be processed by their receiver's (cf. Figure \ref{fig:alertmessage}).

Alert messages contain their sender's identity, alert type and object. Depending on the type, the object is either a sensor and its sensed value leading to the alert or a response to a BFT message about the node itself. How a node can react to a BFT message will be explained in the following section.

%% file: sections/5-architecture.tex
\section{Architecture, Algorithms and Communication Protocol} \label{chap-architecture}

\begin{figure}[h]
  \centering
  \includegraphics[width=1\linewidth]{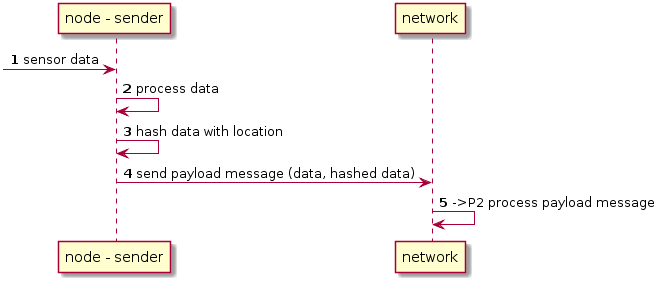}
  \caption{Node sensing a value, preparing and transmitting a payload message (P1).}
  \label{fig:p1}
\end{figure}

The key of the presented concept is the verification of a node's identity by its location and, hence, to trust messages from identified nodes. As established, in addition to payload messages (cf.\ Figure \ref{fig:paymessage}), BFT messages (cf.\ Figure \ref{fig:bftmessage}) are introduced. Figure \ref{fig:p1} depicts a sequence diagram, showing a node generating and transmitting a payload message.

\subsection{Reception of Payload Messages}

\begin{figure}
  \centering
  \includegraphics[width=\columnwidth]{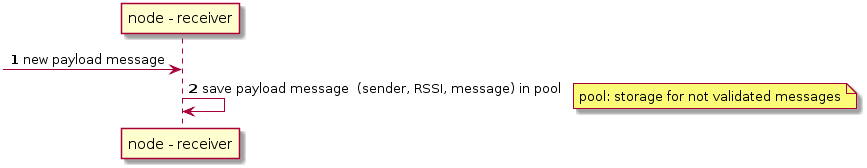}
  \caption{Received payload messages are stored in a pool for later validation (P2).}
  \label{fig:p2}
\end{figure}

The nodes store received messages in a pool of not validated messages (cf.\ Figure \ref{fig:p2}). Messages remain in this pool until the node located its sender and validated that its location is consistent with the location information used to sign the message. It is possible to define a specific time frame during which a message remains in the pool for validation. If validation is not possible during this time frame, the node could discard a message as invalid.

\begin{table*}[h!]
  \centering
  \caption{Topology storage (for sensor node A)}
  \label{tab:topology}
  \begin{tabular}{l|l|l|l|l|l|l|l|l}
    &A&B&C&D&MAC&Sensor&Location&Trust \\
    \hline
    A& & $RSSI_{AB}$ & $RSSI_{AC}$ & $RSSI_{AD}$ & $MAC_A$ & $SensorType_A$ & $Location_A$ & $A_{Tr}$ \\
    B&  $RSSI_{BA}$ & &  $RSSI_{BC}$ &  $RSSI_{BD}$ & $MAC_B$ & $SensorType_B$ & $Location_B$ & $B_{Tr}$ \\
    C& $RSSI_{CA}$ & $RSSI_{CB}$ && $RSSI_{CD}$ & $MAC_C$ & $SensorType_C$ & $Location_C$ & $C_{Tr}$ \\
    D& $RSSI_{DA}$ & $RSSI_{DB}$ & $RSSI_{DC}$ && $MAC_D$ & $SensorType_D$ & $Location_D$ & $D_{Tr}$ \\
  \end{tabular}
\end{table*}

\begin{figure}[ht]
  \centering
  \includegraphics[width=\linewidth]{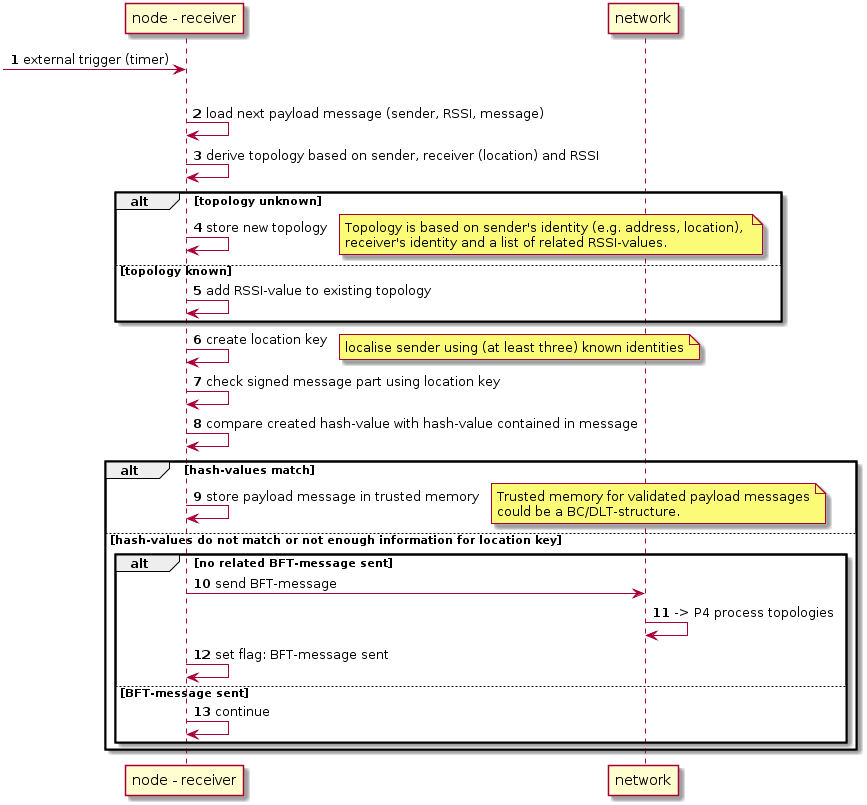}
  \caption{Processing of received transaction messages; extraction of relation and validity check (P3).}
  \label{fig:p3}
\end{figure}

A node's message pool is checked regularly for new or not validated messages (cf.\ Figure \ref{fig:p3}). To locate other nodes and thus validate their messages, specific information for localisation has to be stored. This is done in the topology storage (see Table \ref{tab:topology}). This storage contains information about transmitted and derived locations of nodes and RSSI-values on individual communication paths, e.g.\ in the form $RSSI_{Sender,Receiver} = -42$. The topology storage does not only include topology information for connections from or to the receiving node itself. It also stores topology information of connections between other nodes (limited to those in range). Additionally, the storage has a third dimension, keeping a (limited) history of topology and identity information.

For each message in the pool, topology information is derived and stored (transitions 2 to 5, Figure \ref{fig:p3}). If the extracted connection between two nodes is new to the node, the node adds it to the topology storage. Otherwise, the node adds the RSSI value of the current message to the existing entry. This history of RSSI values is essential. It allows filtering to reduce noise or tracking of other irregularities like node movement.

From stored topology information, a node tries to generate the sending node's location key. It is possible to confirm the generated location key with the received hashed payload by hashing the payload locally (transitions 6 to 8, Figure \ref{fig:p3}). If the two hashed payloads match, the message is successfully validated and will be stored in the receiver's trusted memory (transition 9, Figure \ref{fig:p3}). Otherwise, there is either not enough topology information to localise a node or the known information does not lead to a validation (e.g., in case of movement or manipulation of the sending node). In this case, a BFT message is sent (transition 10, Figure \ref{fig:p3}).

\subsection{Emission of BFT Messages}

\begin{figure}[hb]
  \centering
  \includegraphics[width=1\linewidth]{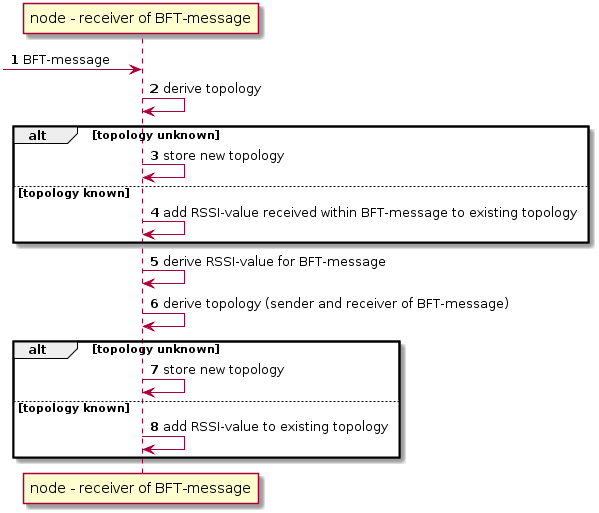}
  \caption{Response to a received BFT message (P4).}
  \label{fig:p4}
\end{figure}

BFT messages ensure the availability of sufficient context information inside the network's nodes to verify the sending node's location. If the context is insufficient or inconsistent with the currently experienced situation, nodes actively try to improve context by sending out BFT messages. They contain the BFT sender's relative location, the payload message sender's identity and its $RSSI_{SR}$. Other nodes react to BFT messages, as depicted in Figure \ref{fig:p4}.

First, the original topology information, i.e.\ the topology information derived by the receiver of the payload message, who is now emitter of the BFT message, is extracted and stored (transitions 2 to 4, Figure \ref{fig:p4}). Then, analogue to processing of payload messages, the relation between sender and receiver of the BFT message is derived and stored (transitions 5 to 8, Figure \ref{fig:p4}).
In practice, the emission of BFT messages is, while possibly triggered by a received payload message, asynchronous to the transmission of payload messages. After a round of BFT messages, nodes in the range should have enough context to decide if a received payload message is to be trusted or not. 

BFT messages of different depth and different triggers for BFT messages are possible. A second-level BFT message could be a reply to a BFT message, triggered by the reception of BFT messages from a non-validated node. This message could include the original BFT message, location and $RSSI_{SR}$, as derived by the BFT message's receiver.  However, the more and the longer BFT messages become, the less light-weight would the protocol be. Therefore this is not suitable for PoL as presented in this paper.

\subsection{Emission of Alerts}

Attack related alerts and \emph{trust scores} inform about the behaviour of nodes that could be an attack. Trust scores are a part of identity in the topology storage (cf. Table \ref{tab:topology}), managed locally by each node. Two conditions determine if the trustworthiness of a node in range is low: A low trust score or the presence of a larger number of BFT messages objecting the node in question. As explained above, BFT messages are emitted when the identity of a node cannot be proven. Multiple BFT messages about the same node, therefore, indicate that something is wrong with that node. The trust score is a second measure and adjusted according to the behaviour of that node.

For a node $A$ (assumed to be trustworthy), suspicious behaviour of a node $B$ would be to receive a BFT message from $B$ about $A$ ($BFT_{BA}$), as long as $A$ is not aware of any movement. There are five reasons for $A$ receiving a $BFT_{BA}$ message:
\begin{enumerate}
    \item Node $A$ has moved,
    \item Node $B$ has moved,
    \item There was a measurement error,
    \item A malicious node $A'$ has sent a payload message, pretending to be $A$ that $B$ responded to, or
    \item Node $B$ is malicious and sends a BFT message to manipulate the network's trust towards $A$.
\end{enumerate}

\begin{figure}
  \centering
  \includegraphics[width=\linewidth]{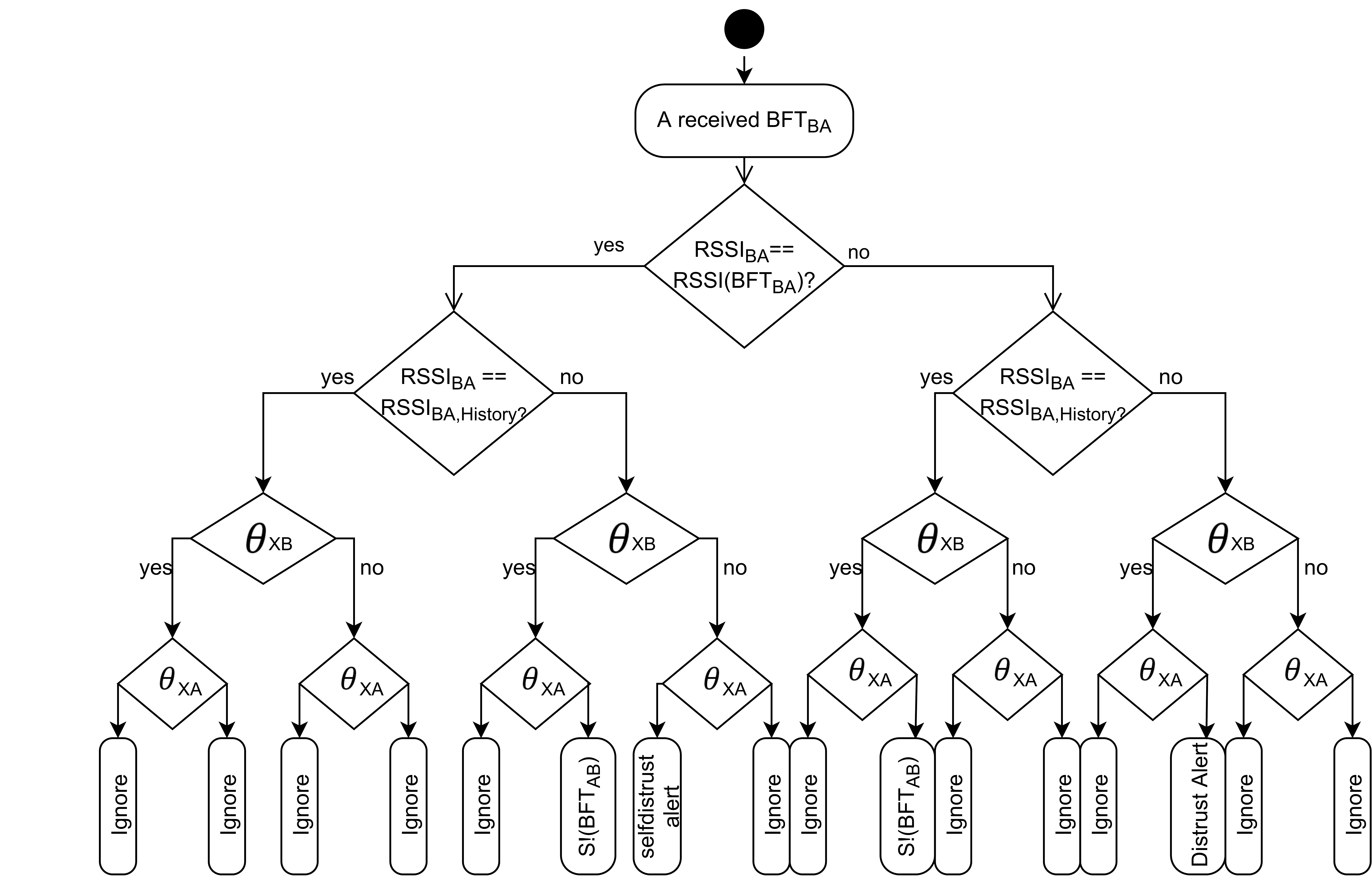}
  \caption{Decision tree depicting how node A processes reception of a BFT message from node B with node A as object.}
  \label{fig:distrust_alert}
\end{figure}

The decision tree in Figure \ref{fig:distrust_alert} depicts, how node $A$ reacts to a received $BFT_{BA}$ message. The tree tends to lead to ``ignore''-states if it is not possible to identify the reason for the BFT message with certainty. This saves operations and keeps the protocol light-weight. In general, the protocol is robust against a node missing a manipulation attempt, especially since it is quite likely that other nodes will identify the attempt anyway, rather than sending a possibly unjustified alert.

Decision levels three and four in the tree in Figure \ref{fig:distrust_alert} are about trustworthiness or rather distrust of one node in another:

\begin{displaymath}
\begin{array}{l}
     \theta_{XA} = (A_{Tr} < \varepsilon) \ \vee \ (|BFT_{
     YA}| > \tau) \big \vert Y \in Node
  \end{array}
\end{displaymath}

Distrust of X against A ($\theta_{XA}$) is a binary function, becoming $true$ if the trust of $X$ in $A$ ($A_{Tr}$) is lower than a limit $\varepsilon$ or $X$ receives a number of BFT messages ($|BFT_{YA}|$) exceeding a limit $\tau$ (dependent on the number of nodes in range). In other words: if $X$ has a low trust score on $A$ or a certain number of nodes in $X$'s range question $A$'s identity, $X$ has distrust in $A$. 

The first two levels of the tree interpret the received BFT message and compare the claimed RSSI value ($RSSI(BFT_{BA})$) with the measured value ($RSSI_{BA}$) and historical values from the topology storage. If all of these values are consistent, the case is ignored completely, as there is no clear indication of why $B$ sent a BFT message. If only the historical values are inconsistent, distrust is taken into account. If $A$ has distrust against $B$, e.g. because several BFT messages with $B$ as an object are received, $B$ might have sent a BFT message because it was the node moved. As a result, $A$ emits a BFT message to indicate a problem with $B$ ($S!(BFT_{AB})$). The problem is not that striking, that an alert has to be sent. However, if history is inconsistent, $A$ has no distrust against $B$, but there is a distrust in $A$ (since $A$ is the perspective, distrust can only mean that several $BFT_{XA}$ messages are received, i.e.\ BFT messages with $A$ as object). Now $A$ must admit that something is wrong with itself. Either $A$ has moved, or an $A'$ tries to steal $A$'s identity. As an honest node, $A$ emits a \emph{``self-distrust'' alert}, meaning that $A$ believes $B$ and thinks that it got attacked or manipulated and thus should not be trusted anymore.

If $RSSI(BFT_{BA})$ claimed by $B$ is inconsistent with $RSSI_{BA}$ measured by $A$, this is suspicious. If $RSSI_{BA}$ is consistent with history, $B$ seems to be at its usual location, so the sent $RSSI(BFT_{BA})$ might be a measurement error. Still, if there is distrust against $B$ (i.e. $B$ has low trust or several nodes seem to see $B$ at a different location) but not itself, $A$ it emits a $BFT_{AB}$ message. If history is inconsistent with $RSSI_{BA}$ and $A$ has distrust against $B$, while no further BFT messages are indicating that $A$ might have been manipulated, $A$ emits a \emph{``distrust'' alert} against $B$, meaning that $A$ believes that $B$ has sent the original BFT message maliciously.

The next step is to implement the reception and processing of a distrust alert. A decision tree analogously to Figure \ref{fig:distrust_alert} becomes very large and is infeasible to depict here. One reason for this is that the receiver (e.g. node $C$) has to ensure that $A$'s alert is not an attack-attempt itself, e.g. by ensuring that the BFT message, the distrust alert replies to, really has been sent. 

In a nutshell, the tree has three outcomes: (1) ignore, (2) accept, and (3) reject. An accept of the distrust alert means, that $C$ also sees $B$ as untrustworthy, as a result, $B$'s trust will be reduced ($B_{Tr}--$). A reject means that $A$'s alert is not trusted and therefore will lead to a reduction of $A$'s trust score and additionally increase the trust score for all nodes that participated in the dissent, i.e. that also indicated a trust problem regarding $A$ by sending a related BFT message:

\begin{displaymath}
\begin{array}{l}
     A_{Tr}--, \forall BFT_{XA}: X_{Tr}++
  \end{array}
\end{displaymath}

To exemplify: A distrust alert by $A$ regarding a BFT message from $B$ is accepted, if 
\begin{itemize}
    \item $C$ has received the related BFT message ($BFT_{BA}$),
    \item $C$ can confirm $A$'s identity, i.e. the measured $RSSI_{AB}$ is consistent with the information in the topology storage and there is no distrust against $A$ ($\theta_{XA}==false$) and 
    \item there is doubt about $B$'s identity, i.e. $RSSI_{BC}$ is inconsistent with information in the topology storage and there is a distrust against $B$ ($\theta_{XB}==true$) -- again, meaning that the trust in $B$ is already low or there is a consent in the network that $B$'s identity has been tampered with, manifested in several $BFT_{XB}$ messages.
\end{itemize}

%% file: sections/6-implementation.tex
\section{Evaluation} \label{chap-implementation}

This section evaluates the utility and practicability of the described concepts. For this purpose, a test scenario has been developed based on two use cases from ongoing research projects. 
The main goal is to show the feasibility of localisation for identification; therefore, the focus lies on the emission of BFT messages implemented and tested on embedded hardware. 

The following describes the test scenario, implementation details and experimental results. Early tests have shown that further effort would be necessary to tackle instability issues of RSSI value measurements. Thus, an analysis tool has been developed to identify a suitable filter and proper parameters. Subsection \ref{subsec: Tool} describes the tool and method.

\subsection{Use Cases \& Test Scenario}
Several ongoing research projects have shown the need for PoL. The two most important use cases that finally inspired the test scenario are:
\begin{itemize}
    \item Freight tracking in logistics as researched in the HANSEBLOC-project\footnote{https://www.hamburg-logistik.net/hansebloc/}. During the transportation of goods, PoL can discover the unwanted and unauthorised movement of goods. Assuming goods transported on pallets, each pallet can be equipped with a SN, thus creating a WSN. Since pallets should not change their relative position during transportation, application of PoL is feasible;
    \item Measurement campaigns in a city's sewage system monitoring as in the research project WaterGridSense4.0\footnote{https://www.watergridsense40.de/}. The deployed WSN consists of one sensor node per surface water drain. Secure supervision of a sewage system to detect illegal external sewage injection is enabled by using PoL. 
\end{itemize} 

These use cases differ in their realisation. They are combined into one experimental test scenario, evaluating the functionality of PoL implementation. For convenience, arbitrary sensor values (temperature) account for the payload. The focus lies on optimising RSSI value filters to enable stable and faultless release of BFT messages, as these are the core of the subsequent PoL security mechanisms. 

\begin{figure}
    \centering
    \includegraphics[width=0.8\columnwidth]{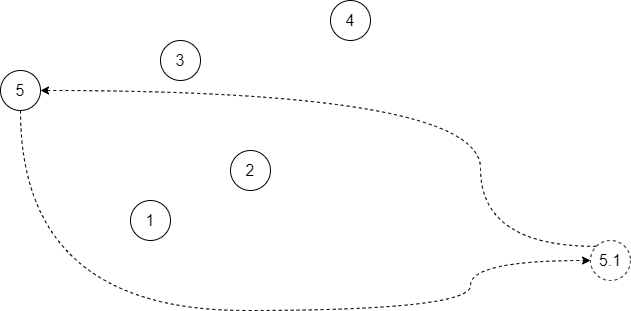}
    \caption{Arrangement of SNs. All nodes static, while node 5 moves from position 5 to 5' and back.}
    \label{fig:sensor movement}
\end{figure}

Five SNs are arranged in a distributed topology on three height levels, as shown in Figure \ref{fig:sensor movement}. Nodes' plane distances vary between a half and two metres. Node 2, 3 and 5 are on the same height level. Node 1 is put one meter below them and node 4 one meter above. While SNs 1 to 4 are static, SN 5 moves twice: (1) it is moved away from the network to position 5' and (2) it is returned into the network to position 5.  

\subsection{Embedded Implementation}
\label{subsec: emb-impl}

Network communication follows the open-source standard \emph{OpenThread}. The five SNs in the experimental scenario are based on Particle Xenon boards equipped with the NRF52840 microcontroller. 
The final hardware developed in the aforementioned research projects uses the same microcontroller on a custom-designed board with a smaller footprint; however, the Particle boards are a suitable prototyping platform.
The Particle Xenon boards offer the possibility to use Zephyr as the operating system and programming with C. 

As mentioned in Section \ref{chap-basics} a SN's identity is defined as the sum of its attributes. For the sake of simplicity, besides a SN's localisation, each SN is identified by its MAC address. Using the MAC address is sufficient for the experimental scenarios. However, real applications should take further attributes into account. The topology matrix is implemented as a struct containing the known RSSI-values. The structs are arranged in a kind of ring storage. Whenever a new RSSI-value is received, either by direct measurement or extracted from a BFT message, it is stored. Yet, there exists no implementation of an initial localisation phase. Therefore, only RSSI-values are available for localisation. BFT messages are implemented and sent when a SN senses a deviation in the RSSI-values. A filter determines the deviation threshold, described in the subsequent subsection. BFT messages can be received and initially handled. Implementation of alerts or responses to BFT messages is still a work in progress.

\subsection{Filter Determination}
\label{subsec: Tool}
Measurement of RSSI-values is very dependent on environmental conditions and affected by noise. The presented concept assumes that for RSSI-values $RSSI_{XA} = RSSI_{AX}$ is true. Initial tests have shown that this is true within a deviation of $\pm 5$dB.
Nevertheless, raw values measured by SNs vary at almost every timestep. Thus, to guarantee a reliable emission of BFT messages, the need for a more sophisticated filter was obvious from the beginning. The main requirements for the filter are:

\begin{itemize}
    \item it shall be able to filter outlier values so that there are no false BFT-messages sent to the network and
    \item it shall not miss actual changes in a node's current location, hence sending too few BFT-messages; 
    \item whilst movement of a SN shall only be indicated by a limited amount of BFT-messages per SN.  
\end{itemize}

The first two restrictions are required so that no false movement is indicated, while actual movements are not dismissed. The third restriction is required to prevent overloading the WSN with redundant messages.  To overcome this problem a two-step process is required -- smoothing RSSI-values and then determine whether to send a BFT-message or not.

\begin{figure}
    \centering
    \includegraphics[width=\columnwidth]{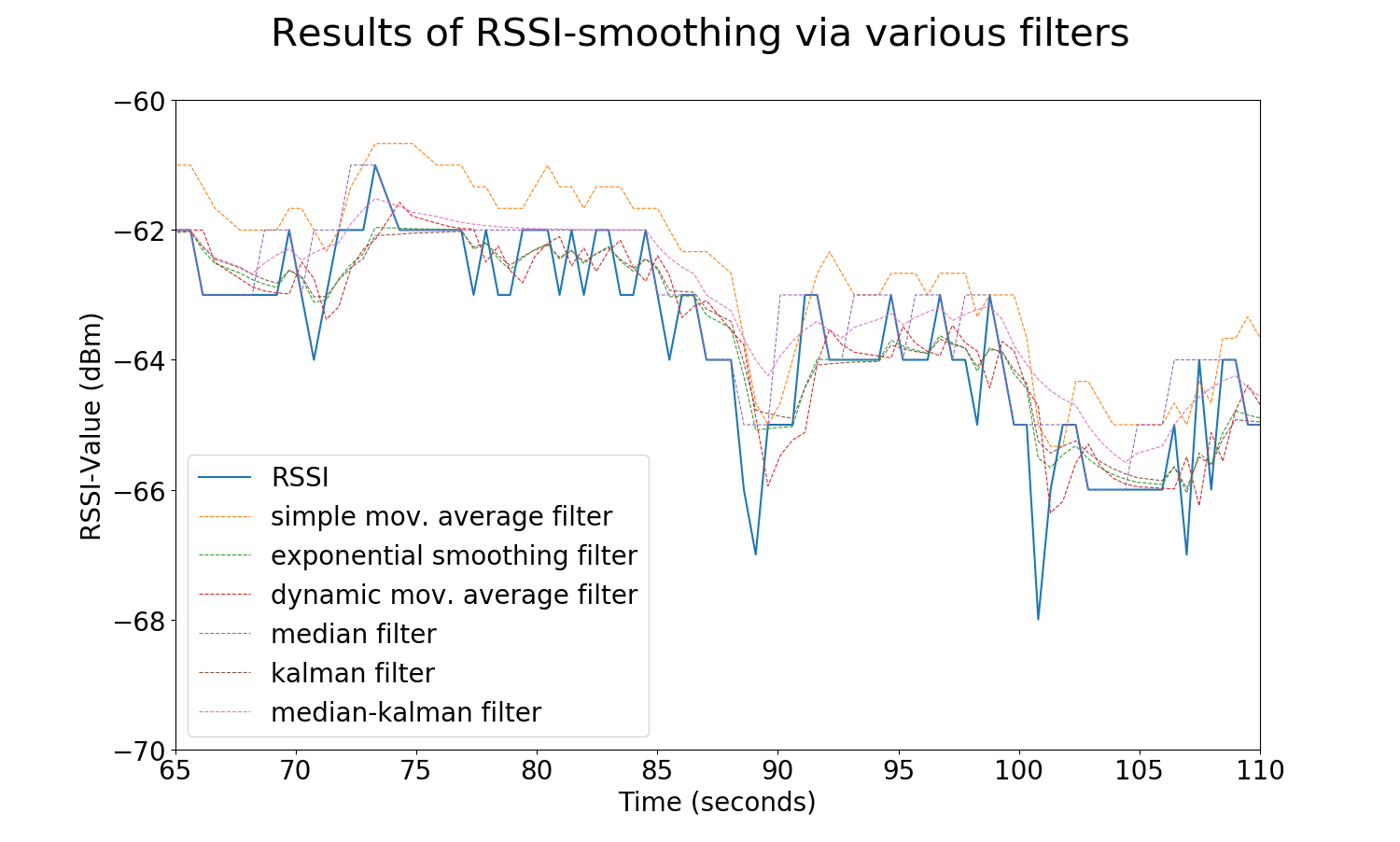}
    \caption{Exemplary close up on RSSI-values and resulting data created by various smoothing filter, showing differences in their smoothing qualitiy. }
    \label{fig:my_label}
\end{figure}

With each experiment, all measured RSSI-values were logged to develop a suitable filter. A self-developed Python-based analysis tool processed the log-files. The analysis tool provides various filters: simple moving average filter, exponential smoothing filter, dynamic moving average filter, Gaussian filter, median filter, and Kalman filter (c.f. \cite{indoor-localisation}). Figure \ref{fig:my_label} shows the varying smoothing quality of the different filters. By further examination, a combination of median filter and Kalman filter has proven to fit best. Thus raw measurement values are now first fed into a median filter and, subsequently, the result values are handed to a Kalman filter. 
The second process step is a function to trigger BFT-messages. The function uses the smoothed RSSI-values and triggers a BFT-message if the current values differ from the previous values by more than a previously set threshold.

\subsection{Experimental Results}
In the experimental setup, the MAC-address of the SNs is used as a reference. This is the main identifier of each SN. This identification shall be backed up by the node's location, based on RSSI values relative to the other nodes in range. Location is the ``true identity'', as described in the concept in Section \ref{chap-concept}. A manipulation is assumed when the MAC-address is not backed up by the location anymore, i.e. when a SN is moved or a malicious node tries to fake another SN's MAC-address but sending from a different position. In this case, it is expected that (a majority of) other SNs in the network recognise the discrepancy between MAC-address and location and therefore emit a BFT-message. 

\begin{figure*}
    \centering
    \includegraphics[width=\textwidth]{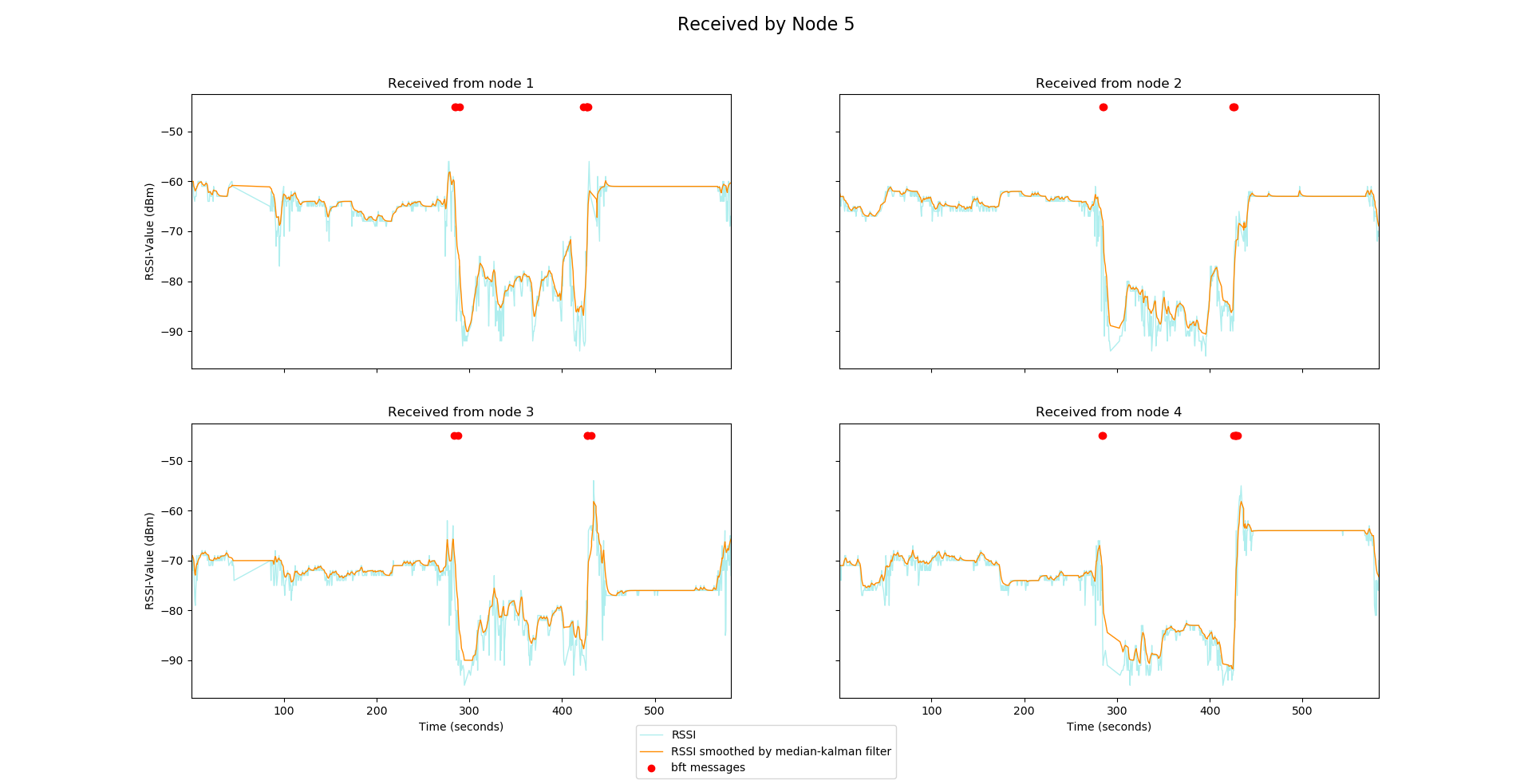}
    \caption{RSSI-values (blue) and BFT-messages (red) received by SN five during experimental setup and median-kalman filter smoothed RSSI-values (orange).}
    \label{fig:data-analysis}
\end{figure*}

The experiment shall prove that a movement is recognised, while false positives are minimised. Correct emission of BFT messages is the fundamental requirement to implement and utilise PoL. While a couple of test runs have shown that RSSI-measurement is symmetrical within a range of 5 dB, further runs have been performed to gather data, to find useful filter parameters as described before. Figure \ref{fig:data-analysis} depicts the data collected through measurement with the SN arrangement depicted in Figure \ref{fig:sensor movement} using the SNs described in \ref{subsec: emb-impl}, using the final filter settings. It shows the raw RSSI-values that SN five received from the other SNs. Also shown are the RSSI-values smoothed by the defined filter-combination of a median and a Kalman filter.
The change of position relative to the other four nodes can be seen in the RSSI-values node five determined about these nodes. As it was a movement away from the other nodes, the depicted RSSI-values are decreasing (at $\sim$300 seconds). Then staying constant for the time the SN held its distant position and increasing again as it was put back to its original position.
Thus, because the chosen filter smooths outliers while movements are still recognisable, it is suitable for the desired use cases. 

The figure also shows that sensor node five received BFT-messages about itself from each other node. These were received, and hence previously sent, directly after sensor node five moved. As figure \ref{fig:data-analysis} depicts, the movements of sensor node five triggered up to two BFT-messages at each sensor node. 

Given these results, it can be stated that the selected filter is suitable for the test scenario and that the basic mechanics of PoL are implemented and tested successfully:
\begin{itemize}
    \item RSSI-values can be measured and used to gain a stable image of the nodes' location in the network, i.e. the network topology
    \item Based on the RSSI-values, nodes can identify their peers
    \item Manipulation of a node, resulting in transmissions from a different location, can be recognised by its peers
    \item BFT messages are emitted and received reliably, allowing the nodes to communicate detected manipulation.
\end{itemize}

In consequence, the basic requirements for more comprehensive implementation and evaluation of PoL are met.

%% file: sections/7-conclusion.tex
\section{Conclusion} \label{chap-conclusion}

\subsection{Summary}

Proof-of-location (PoL) is a new approach to enable secure communication in IoT. It aims to identify maliciously or manipulated network participants based on those devices' location. PoL uses a new type of communication messages, so-called BFT messages, to identify nodes and manipulations.  With BFT messages, nodes can reach a consensus -- the PoL -- on the nodes' trustworthiness inside the network. This paper presented PoL in detail and an approach for its implementation. The first steps to develop PoL into a protocol have been taken. Central parts of PoL have been implemented on prototypical IoT-hardware, communicating via Bluetooth low energy. The implementation identifies nodes based on RSSI-values and shows PoL's usefulness and feasibility as a light-weight protocol. Several experiments have led to an adequate filter function, delivering reliable results using PoL. Emission of BFT-messages in case of identification of manipulation attempts has been implemented and successfully proven.

\subsection{Future Work}

The results presented in this paper show the fundamental feasibility of PoL. Future research includes the complete implementation and formal verification of the remaining communication processes described above. Further processing of BFT and alert messages still needs to be implemented and evaluated. Finally, an automatic adjustment, depending on environmental conditions, of smoothing parameters in the filter function is desired. In this way, the approach could adapt more easily to different environments.